# Does Environmental Attention by Governments Promote Carbon Reductions?

Yichuan Tian [1,*]

[1] School of Law, Anhui Normal University, Wuhu 241000, China
\* Correspondence: tianyichuan@ahnu.edu.cn

## Abstract

The carbon-reducing effect of attention is scarcer than that of material resources, and when the government focuses its attention on the environment, resources will be allocated in a direction that is conducive to reducing carbon. Using panel data from 30 Chinese provinces from 2007 to 2019, this study revealed the impact of governments' environmental attention on carbon emissions and the synergistic mechanism between governments' environmental attention and informatization level. The findings suggested that (1) the environmental attention index of local governments in China showed an overall fluctuating upward trend; (2) governments' environ-mental attention had the effect of reducing carbon emissions; (3) the emission-reducing effect of governments' environmental attention is more significant in the western region but not in the central and eastern regions; (4) informatization level plays a positive moderating role in the relationship between governments' environmental attention and carbon emissions; (5) there is a significant threshold effect on the carbon reduction effect of governments' environmental attention. Based on the findings, this study proposed policy implications from the perspectives of promoting the sustainable enhancement of environmental attention, bringing institutional functions into play, emphasizing the ecological benefits and strengthening the disclosure of information.

**Keywords** governments' environmental attention; carbon reduction; informatization level; local government

## 1 Introduction

Since the implementation of the reform and opening-up policy, China's sustained rapid economic growth has created one of the great miracles in the history of the world's economic development. However, while China has made outstanding achievements in the area of economic development, it is also facing enormous international pressure to reduce carbon emissions. In 2007, China exceeded the US as the world's largest carbon emitter(Zhang et al. 2018). In 2020, the Chinese mainland accounted for 30.7% of global CO2 emissions, compared with 13.8% in the US and 11.1% in Europe (BP 2021). This shows that China's carbon emissions are relatively high and that reducing these emissions is an essential requirement for the sustainable development of the Chinese nation. It is for this reason that China solemnly pledged at the 75th session of the UN General Assembly to take stronger policy measures to reach peak carbon dioxide emissions by 2030 and achieve carbon neutrality by 2060. Under the guidance of the concept of harmonious coexistence between humans and nature, the way to promote reductions in carbon emissions in an orderly manner has become a key issue of academic concern, not only regarding the green transformation of economic development, but also the sustainable development of the Chinese nation.

Carbon-reducing strategies can be divided into two forms: market-driven and government-driven (Xu et al. 2023). On the one hand, market-driven reduction strategies focus on carbon pricing mechanisms such as carbon taxes (Hájek et al. 2019) and national emissions trading schemes (Heggelund et al. 2019). Taking the national emissions trading scheme as an example, although China's carbon trading market system is gradually improving and making some contributions to reducing carbon emissions, it still faces challenges such as insufficient domestic demand, limited financial participation, an inadequate regulatory infrastructure, and excessive government intervention (Lo 2016), which also suggests that relying on the market alone to achieve carbon emissions is not enough. On the other hand, in the current academic discussion of government-based strategies for reducing emissions, scholars have mostly focused on the intervention of the "visible hand" of fiscal environmental expenditure on carbon emissions (Wang &Li 2019, Xie et al. 2023), but have often overlooked the possible role of governments' environmental attention (hereafter referred to as GEA) in reducing carbon emissions.

As a topic that has been emerging in the field of environmental policy research in recent years, the emergence and development of GEA research in China have gone through three stages. The first was the nascent period, during which research on government attention had not yet been extended to the environmental field, though the accumulation of knowledge related to GEA by relevant research should not be overlooked. For example, Wen (2014) measured the central government's attention to basic public services by drawing on the working reports of the Chinese central government over the period 1954-2013, which also provided lessons to be drawn from for the measurement of GEA. The second was the development period, during which scholars began to adopt word frequency in policy documents as a proxy variable for the governments' environmental regulations (Chen et al. 2018); however, probably due to the lack of academic precedents, most studies did not combine this innovation in the empirical approach with attention theory. The third was the mature period, when the theory and empirical analysis of GEA began to be effectively integrated. For example, Bao andLiu (2022) quantified the GEA of local governments in China from 2014-2019 on the basis of governmental working reports from 286 prefecture-level cities in China, and examined the relationship between GEA and air pollution control. Liu et al. (2023) found that GEA made a significant contribution to reductions in regional environmental pollution and emissions. It should be noted that while scholars have begun to explore the important role of GEA in environmental governance, there is still a gap in research on how GEA affects carbon emissions, which leads to the first research question of this study. Can GEA contribute to reductions in carbon? At the same time, China is undergoing an unprecedented digital transformation, with the number of internet users growing from 59.1 million in 2000 to 1.032 billion by 2021, and the number of internet broadband users growing from 3.253 million in 2002 to 536 million by 2021 (Pan et al. 2023). In academia, numerous studies have also demonstrated that economic development of the internet, internet infrastructure, and ICT can significantly improve performance in terms of carbon emissions (Haini 2021, Kou &Xu 2022, Wang et al. 2022). Thus, the second question arises. Does the informatization level play a moderating role between GEA and carbon emissions?

To answer these two research questions, this study constructed a GEA index based on the panel data of 30 Chinese provinces from 2007-2019 using Python's loop statement and the Jieba library. The data were subjected to two-way fixed effects (TWFE) and two-stage least squares (2SLS) regressions, providing empirical evidence from China to clarify the relationships among governments' environmental attention, informatization level, and carbon emissions.

## 2 Literature Review

### 2.1 Governments' Environmental Attention: Concept, Theory, and Measurement

The concept of attention first came from early psychological research, with James (1890) pioneering the idea that attention is the process by which consciousness, in a clear and rapid form, selects one of many possible objects or ideas, and that the essential feature of attention is the focus and concentration of consciousness. Lauwereyns (2011) further used the preferred mechanism to refine the concept of attention, arguing that the existence of preferences makes attention sequential and that attention is always used to describe the priority that should be given to an event, thereby discarding other events. In the mid to late 20th century, scholars such as Simon and March began to introduce the concept of attention into organizational behavior. According to Simon (1997), attention in organizational science refers to the process by which managers selectively focus on certain information while ignoring other parts. March (1994) focused on how attention affects managers' decision-making, stressing that managers' decisions are largely dependent on how effectively they can allocate their attention. On the basis of Simon's research, Jones extended the concept of attention to the field of governmental decision-making. For Jones (1994), the concept of attention implied selectivity and a decision-making mechanism. On the one hand, governmental decision-makers target a particular aspect of the social environment as a result of an attention-driven process; on the other hand, through this selection, a prominent feature of the social environment is brought into the structure of decision-making and becomes the object of the decision maker's choice. Through a review of previous literature, this study attempted to define governments' environmental attention as follows: governments' environmental attention refers to the intensity of attention that governmental decision-makers selectively give to environmental issues, which influences their preferences for making and implementing policies.

Currently, there are two main theories of governments' environmental attention: the attention-based view proposed by Ocasio (1997), and the attention-driven policy choice model proposed by Jones andBaumgartner (2004). In the theoretical system of the attention-based view, the attentional resources of governmental decision-makers are as limited as rationality, so decision-makers have to allocate their limited attentional resources unequally to various public issues and solutions. This process of distribution consists of three specific steps: attention, interpretation, and action. Specifically, attention means that governmental decision-makers selectively focus on their preferred public issues and the related information before entering the decision-making process. Secondly, in the course of the decision-making process, governmental decision-makers explain their concerns to make their individual values and preferences seem more reasonable and feasible, thus leading to a collective consensus. Finally, governmental decision-makers formulate solutions to public issues and promote their implementations according to their attention and interpretation. At one level, this process of allocating attention is essentially a process of communication and consensus leading to collective action, and its use in analysis needs to account for the influence of contextual factors in decision-making, which means that the attention-based view is strongly embedded. The

'attention-driven policy choice model', on the other hand, sees attention more as a symbol that can be progressively concretized in the operation of the hierarchical structure as a selective allocation of government resources to certain public policies. When the attention of policy makers shifts, government policies change with it (Liu et al. 2023).

Similar to tracking people's gaze, tracking language use can reveal what people are paying attention to; in other words, the categories of words used in a text can reveal the areas of focus of attention (Tausczik &Pennebaker 2010). In China, the governmental working report is the most important official document prepared by governments at all levels to summarize the economic and social achievements of the past year, and to establish working plans and specific targets for the coming period (Chen et al. 2018). Therefore, this study chose to use the governmental working reports of China's provincial governments as a data source and measured the environmental attention of local governments based on the word frequency (as a percentage) of environmental keywords in the reports. Firstly, in line with the studies of Bao andLiu (2022), Liu et al. (2023), and a close reading of the 390 selected governmental working reports, three metrics of environmental protection, pollution control, and ecological civilization were summarized, and the keywords under each specific metric are detailed in Table 1. Secondly, the tasks of separating the words and determining the word frequency statistics of the governmental working reports were completed with the help of Python loop statements and the Jieba library. Finally, data cleaning and table merging operations were carried out on the constructed indicators of governments' environmental attention via Excel to facilitate subsequent of building a model and validation of the hypotheses.

**Table 1.** Dimensions and Keywords

| Dimensions in Chinese | Dimensions in English | Keywords in Chinese | Keywords in English |
|---|---|---|---|
| 环境保护 | Environmental protection | 环保 | Environmental protection |
| | | 绿化 | Greening |
| | | 环境质量 | Environmental quality |
| | | 空气质量 | Air quality |
| | | 环境治理 | Environmental governance |
| 污染治理 | Pollution control | 治污 | Pollution control |
| | | 低碳 | Low carbon |
| | | 节能 | Energy-saving |
| | | 减排 | Emission reduction |
| | | PM2.5 | PM2.5 |
| | | 二氧化硫 | Sulfur dioxide |
| | | 二氧化碳 | Carbon dioxide |
| | | 扬尘 | Dust |
| | | 新能源 | New energy |
| 生态文明 | Ecological civilization | 生态 | Ecology |
| | | 绿水青山 | Clear waters and green mountains |
| | | 自然保护区 | Nature reserve |

## 2.2 Governments' Environmental Attention and Reductions in Carbon

It was clear from the process of establishing the theoretical foundations that, compared with material resources, attention is not only limited but also scarce. As decision-makers, local government officials cannot divide their attention evenly among the various public issues but need to make judgments about the priority of public issues and solutions. Therefore, when the government focuses its attention on the environment, it will inevitably lead to a shift in the allocation of public resources towards environmental governance efforts such as reductions in carbon. This is both because attention is a highly condensed version of the subjective will of decision-makers and because attention itself represents a precursory signal of politics. A higher level of environmental attention by the government means that the decision-makers and their demonstrated willingness to govern the environment will be stronger, leading to more aggressive action to reduce carbon emissions. Similar to entrepreneurs, the policies pursued by local governmental decision-makers change in response to changes in attention. From the perspective of attention, governments' environmental attention can drive the formation of motivations and the behaviors of policies. The selectivity of government decisions is gradually transitioned in favor of environmental governance through the selection mechanism of environmental attention, thus enabling environmental policy to be allocated more public resources.

Let us now return to the Chinese context. "The Great Divergence" has been a popular topic of economic historiography in recent years, and a new view of history now suggests that the root cause of this phenomenon was the failure of ancient China to develop industrial capitalism despite its historically developed market economy (Li 2016). Further, this was due to the difference between the Chinese "Confucian-legalist state" and the European "Corporate state". The kernel of state legitimacy in the Confucian-legalist state expressed by the concept of the "Mandate of Heaven", which states that rulers have the divine right to rule, but only if they take care of the wellbeing of their people; if they fail to do so, they risk being overthrown (Zhao 2015). As Chinese modernization progressed, so too did the "Mandate of Heaven" concept. As the Chinese government put it, "The Chinese path to modernization is a process of modernization of harmony between humanity and nature", and the continued promotion of peak carbon and carbon neutrality is a requirement for achieving harmony between humanity and nature. This modern transformation of the state's legitimacy also applies to Europe and the United States, where for almost 50 years, Rawls' second principle of justice has not only reflected his preference for the least advantaged (Rawls 1999), but also provided the theoretical foundation for the gender justice and racial justice movements. Similarly, justice is also taking shape in the environmental field, one of the elements of which is the need for developed countries to take more responsibility for reducing carbon. Here, we present a discussion of why governments' environmental attention has a carbon-reducing effect at the level of political culture.

China's distinctive administrative system is also driving environmental attention to creating a carbon-reducing effect. Unlike other post-modernization countries, China has the unique institutional framework of a unitary system, which is characterized by a centralized, sovereign power held by a single state body, with local governments required to act at the will of the central government. At the same time, the division of competencies between the central and local state institutions gives local governments the space to exercise their autonomy. This dialectical unity of centralization and decentralization constitutes the interlocking relationship between the central and local governments in China, known as—the "Tiao-Kuai relationship". "Kuai" refers to the vertical hierarchy from the central to the local government, while "Tiao" is the horizontal system of functions based on administrative divisions. This provides realistic grounds for the promotion championship model, where, for a long time, the Chinese central government has typically used the control of personnel to incentivize local officials to promote economic growth (Li &Zhou 2005). Nowadays, with the Chinese government making the construction of ecological civilization an important part of "The Five-Sphere Integrated Plan", the "competition" in the promotion championship model is gradually shifting from a single GDP competition to a diversified competition. When the level of carbon emissions becomes an important indicator in performance appraisals and the competition for jobs, local officials will see environmental attention of their superiors as an indicator of incentive and will consciously shift their focus and allocate resources towards reducing carbon. It is important to note, however, that when the governments' attention is too focused, it may also give rise to a 'blame avoidance mechanism' in the implementation of local government policies. Accordingly, the following hypothesis was proposed:

**H1:** Local governments' environmental attention has a negative impact on carbon emissions.

**H2:** The impact of local governments' environmental attention on carbon emissions is likely to be non-linear.

## 2.3 The Moderating Effect of the Level of Informatization Technology

Within the framework of bounded rationality, Simon suggests that information is not a factor of scarcity; what is scarce is attention (Simon 1971). However, information overload has become a common problem in today's society, and because of bounded rationality, people are only able to process a fraction of this information. This is where the value of information processing power comes to the fore. In other words, it is true that in the era of big data, information is not a scarce factor, but the ability to process information is as scarce as attention. On the other hand, the operation of public power is characterized by its concealment. In the government's decision-making process, only the input of public issues and the output of public policies are visible to the public, while the formulation, adoption, and approval of policies are hidden in a black box that is opaque to the public and outside observers (Doshmangir et al. 2015). The context of the information age and the citizens' demands to open the black box of policies are also driving innovations in government departments. Over the past two decades or so, governments have used information and communications technology (ICT) to integrate their internal functions and improve their delivery of services, a trend that scholars have termed e-government (Manoharan &Ingrams 2018). On the one hand, a high level of information technology facilitates public participation, expert discussions, and media attention, which, in turn, provides a constant stimulus for governmental decision-makers to maintain a strong focus on environmental issues (Wang &Li 2017). On the other hand, the government information disclosure index, an important indicator of the performance of e-government (Wu &Guo 2015), is inextricably linked to the level of development of local informatization. At the same time, the Environmental Attention Index, based on governments' working reports, reflects the preferences of decision-makers at the outset of formulating a policy and anticipates the potential benefits of the policy, a process that is undoubtedly facilitated by the informatization level. Accordingly, the following hypothesis was proposed:

**H3:** The informatization level plays a positive moderating role in the influence of local governments' environmental attention on carbon emissions.

## 3 Materials and Methods

### 3.1 Data sources

Considering the availability of data, 30 provinces in China (excluding Tibet, Taiwan, Hong Kong, and Macao) were selected as the sample for analysis, with data spanning the period 2007-2019, with which the carbon-reducing effect of government attention was examined. The data for the control and moderating variables were obtained from the National Bureau of Statistics of China, the China's environment yearbooks, and provincial statistical yearbooks. The dependent variable, carbon emissions per capita, was estimated mainly through the carbon emission factors provided by the IPCC (2006), and the independent variable, GEA, was derived from a textual analysis of the governmental working reports from the 30 provinces mentioned above.

### 3.2 Description of the Variable

#### 3.2.1 Dependent variable: carbon emissions per capita

The dependent variable in this study was carbon emissions per capita. As carbon emissions are not currently published directly by region, this study chose to use the carbon emission factors provided by the IPCC (2006) to estimate them, based on the final consumption of eight fossil energy sources: coal, coke, crude oil, gasoline, paraffin, diesel, fuel oil, and natural gas. The formula is shown below:

$$CEF_i = H_i \times CH_i \times COR_i \times \frac{44}{12} \times 10^{-6} \tag{1}$$

$$C = \sum_{i=1}^{8} E_i \times CEF_i \tag{2}$$

In equation (1), $H_i$、$CH_i$、$COR_i$ are the average low calorific value, the carbon content per unit of calorific value, and the rate of carbon oxidation, respectively. The carbon emission factor $CEF_i$ for fossil energy $i$ can be derived by the formula, which is then substituted into equation (2), and the regional carbon emissions $C$ can be obtained from the energy consumption $E_i$ provided by the National Bureau of Statistics of China. On this basis, the carbon emissions of each region were divided by the total population of the region to finally obtain the carbon emissions per capita.

#### 3.2.2 Independent variable: governments' environment attention (GEA)

The independent variable in this study was governments' environmental attention. As mentioned earlier, the governmental working reports are the good reflection of the scope of the government attention over a period. Therefore, by systematically sorting through 390 working reports issued by 30 local governments in China between 2007 and 2019, as well as drawing on the keywords and definitions of the governments' environmental attention in relevant studies, this study finally chose to start from the three dimensions of environmental protection, pollution control, and ecological civilization, and extracted 21 environmental keywords. Based on this, the metrics of GEA were constructed by using Python loop statements and the Jieba library to count the word frequency of the extracted keywords and dividing the number of keyword words' frequencies by the total number of words in the governmental working reports.

### 3.2.3 Moderating variable and Control variables

For the moderating variable, the informatization level, this paper uses the total postal and telecommunications business as a percentage of regional GDP. Meanwhile, to minimize the interference of confounding factors on the estimation of main effects, this paper selects energy consumption (Alshehry &Belloumi 2015), industrial structure (Zhu 2022), the number of environmental administrative penalty cases (Liu et al. 2022), fiscal environmental protection expenditure (Xu et al. 2023), foreign direct investment, and environmental regulation intensity (Zhang et al. 2020) as control variables by referring to the existing related literature. The descriptive statistics for each variable are shown in Table 2.

**Table 2. Summary statistics**

| Variable | Measurement | Obs | Mean | Std | Min | Max |
| --- | --- | --- | --- | --- | --- | --- |
| CO2 | Total carbon emissions / population | 390 | 10.151 | 7.054 | 2.627 | 43.601 |
| GEA | Keyword frequency/text frequency | 390 | 0.592 | 0.208 | 0.122 | 1.405 |
| IT | Total postal and telecommunications services/GDP | 390 | 0.059 | 0.038 | 0.014 | 0.236 |
| ER | Investment in industrial pollution control completed / Value added of secondary industry | 390 | 0.003 | 0.003 | 0.000 | 0.025 |
| ENE | Electricity consumption/national total | 390 | 0.033 | 0.023 | 0.003 | 0.104 |
| INDU | Tertiary sector output / Secondary sector output | 390 | 1.082 | 0.622 | 0.500 | 5.169 |
| FDI | Amount of foreign direct investment/GDP | 390 | 0.022 | 0.017 | 0.000 | 0.082 |
| LNPE | The number of environmental administrative penalty cases received (Ln.) | 390 | 7.399 | 1.825 | 0.000 | 10.713 |
| LNEXP | Financial environmental protection expenditure (Ln.) | 390 | 4.482 | 0.781 | 1.671 | 6.617 |

### 3.3 Establishment of the Model

To test the effect of GEA on carbon emissions per capita, the following benchmark regression model was constructed:

$$Y_{i,t} = \beta_0 + \beta_1 GEA_{i,t} + \beta Controls_{i,t} + \mu_i + \lambda_t + \varepsilon_{i,t} \quad (3)$$

In equation (3), $Y_{i,t}$ denotes the carbon emissions per capita of province $i$ in year $t$, $GEA_{i,t}$ is the independent variable of governments' environmental attention, $Controls_{i,t}$ is a set of control variables, $\beta_0$ is the intercept term, $\beta 、 \beta_1$ denote the parameters to be estimated and $\varepsilon_{i,t}$ is the random disturbance term. Given the existence of fixed effects across provinces and time, province-fixed effects $\mu_i$ that do not change over time and year-fixed effects $\lambda_t$ that do not change over time were added.

To eliminate estimation errors arising from the possible two-way causality between GEA and carbon emissions per capita, a two-stage least squares (2SLS) regression model was constructed:

$$GEA_{i,t} = \alpha_0 + \alpha_1 GEA_{i,t-1} + \beta Controls_{i,t} + \mu_i + \lambda_t + \varepsilon_{i,t} \rightarrow \widehat{GEA_{i,t}} \quad (4)$$

$$Y_{i,t} = \beta_0 + \beta_1 \widehat{GEA_{i,t}} + \beta Controls_{i,t} + \mu_i + \lambda_t + \varepsilon_{i,t} \quad (5)$$

In the first stage regression of 2SLS, the instrumental variable $GEA_{i,t-1}$, which satisfies the conditions of both correlation and exogeneity, could strip out the exogenous part of the endogenous variable $GEA_{i,t}$ to obtain the fitted value $\widehat{GEA_{i,t}}$. The second stage of the regression then substituted the fitted values thus obtained into the baseline model for instrumental variable regression, thereby obtaining a consistent estimator of the coefficient $\beta_1$ of the endogenous variables $GEA_{i,t}$.

To test the moderating role of informatization level in the relationship between GEA and carbon emissions per capita, an interaction term model that introduced the moderating variables was constructed:

$$Y_{i,t} = \beta_0 + \beta_1 GEA_{i,t} + \beta_2 C\_GEA_{i,t} \times C\_IT_{i,t} + \beta_3 IT_{i,t} + \beta Controls_{i,t} + \mu_i + \lambda_t + \varepsilon_{i,t} \quad (6)$$

Considering the potential problem of multicollinearity between the interaction term and the original variables, this study chose to center the GEA on the informatization level. It should be noted that this centralization does not affect the reliability of the results of the test of a moderating effect, but it does lead to the fact that it only reduces the problem of non-essential multicollinearity(Dalal &Zickar 2012). On this basis, the centralized variables were cross-multiplied to finally obtain $C\_GEA_{i,t} \times C\_IT_{i,t}$. When the coefficient of $\beta_3$ was significant, this indicated that the moderating effect of the informatization level was significant. If the coefficients of $\beta_2$ and $\beta_1$ were in the same direction, then there was a positive moderating effect; if not, then there was a negative moderating effect. The remaining variable symbols in equation (6) are equivalent to those in equation (3), and the regression models presented above were estimated using robust standard errors.

To test whether the carbon reduction effect of GEA is non-linear, the panel threshold model was used (Hansen 1999). The panel threshold regression model set for the specific context of this study is as follows:

$$Y_{i,t} = \beta_0 + \beta_1 GEA_{i,t} \times I(GEA_{i,t} \leq \theta_1) + \beta_2 GEA_{i,t} \times I(\theta_1 \leq GEA_{i,t} \leq \theta_2) + \ldots$$
$$+ \beta_{n+1} GEA_{i,t} \times I(GEA_{i,t} > \theta_n) + \beta Controls_{i,t} + \mu_i + \lambda_t + \varepsilon_{i,t} \quad (7)$$

where $I(\cdot)$ is the characteristic function, $\theta$ is the unknown threshold, and the rest of the variable symbols are consistent with the above.

## 4 Results and Discussion

### 4.1 Governments' environmental attention (GEA) in China

Figure 1 and Figure 2, respectively, show the trends and regional differences in the average GEA index of local governments in China from 2007 to 2019. It can be seen that the GEA of local governments in China as a whole showed a fluctuating upward trend, indicating that the environment is gradually becoming an important factor influencing governmental decisions and officials' behavior. At the same time, the overall increase was not significant, reflecting that although the government has made some progress in environmental governance, there is still more room to improve the importance it attaches to the environment compared with the sectors of economic and public health. From the perspective of regional differences, the provinces with a high GEA index included both the more economically developed Fujian and the less economically developed Qinghai; the provinces with a low GEA index included both the more economically developed Shanghai and the less economically developed Xinjiang, indicating that there was no significant correlation between GEA and a regional level of economic development.

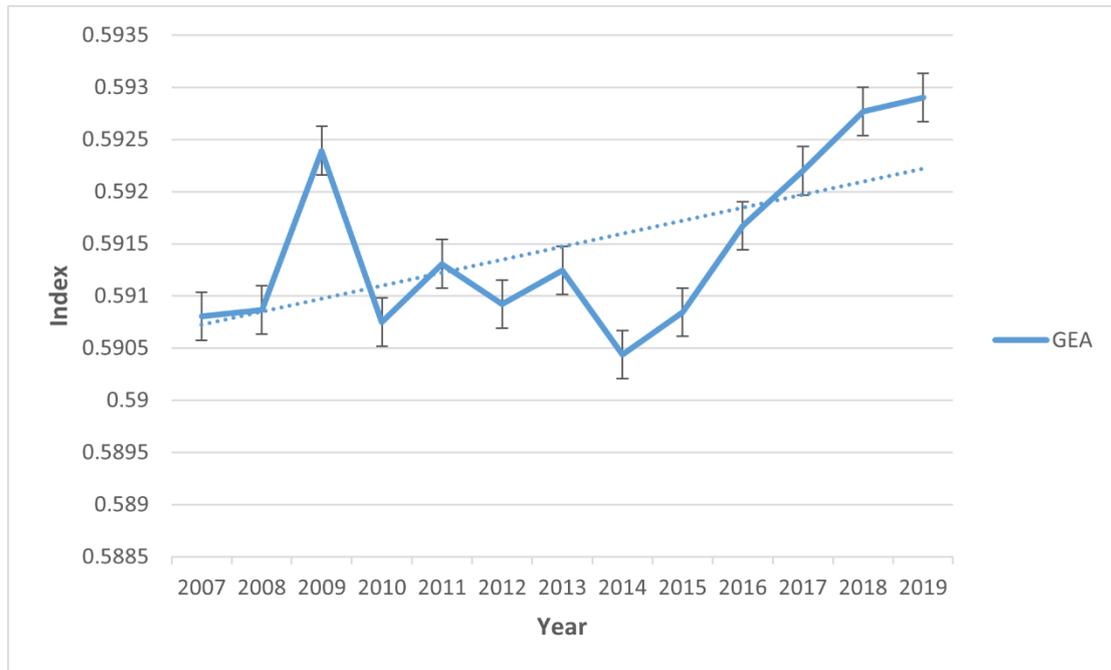

**Figure 1. Trends in the average GEA index**

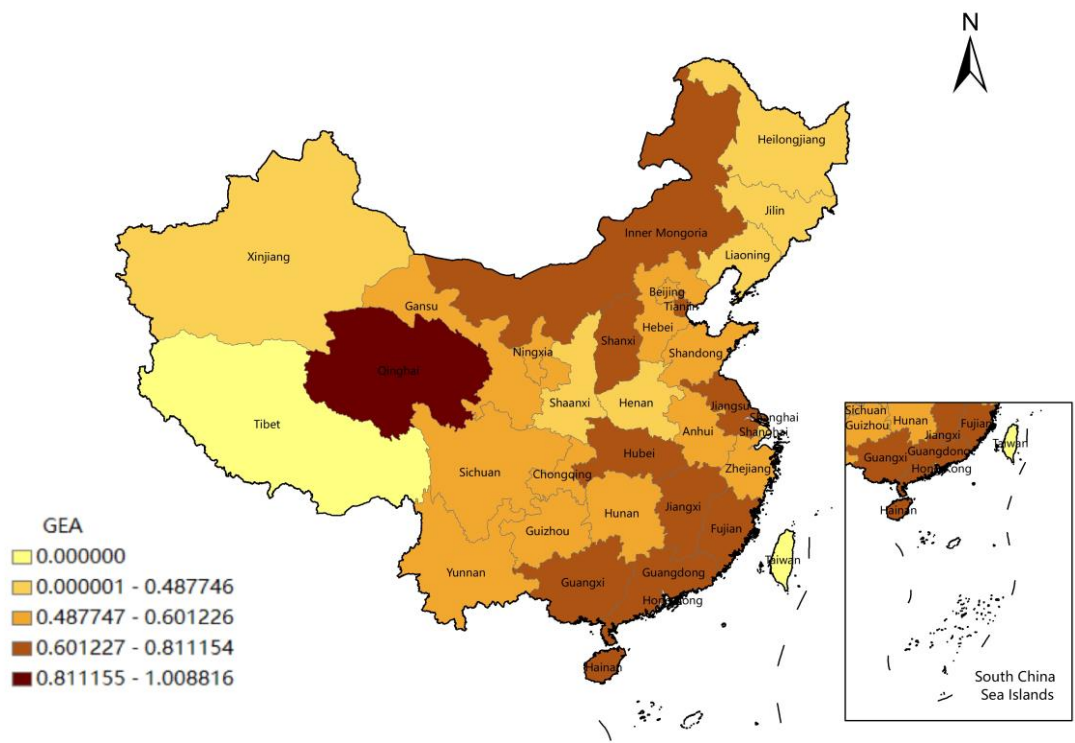

**Figure 2. Regional variations in the average GEA index**

## 4.2 Baseline regression analysis

The degree of multicollinearity was tested by analyzing the correlation between the explanatory variables before constructing the regression model. Table 3 shows that the correlation coefficients between most of the explanatory variables were below 0.3, indicating that there was no serious problem of multicollinearity. Moreover, to determine whether a random effect or a fixed effect should be chosen, a Hausman test was conducted in this study, which rejected the original hypothesis that the random term was uncorrelated with the explanatory variables; therefore, a fixed-effects model was used. Considering that the results of estimation from the random effects model still have some reference value, they have also been presented. Table 4 presents the results of the benchmark regressions, where (1) is a two-way random effects model and (2) is a two-way fixed effects model.

**Table 3. Statistical correlations**

|       | GEA    | IT     | ER     | ENE    | INDU   | FDI    | LNPE   | LNEXP |
|-------|--------|--------|--------|--------|--------|--------|--------|-------|
| GEA   | 1      |        |        |        |        |        |        |       |
| IT    | -0.163 | 1      |        |        |        |        |        |       |
| ER    | 0.005  | -0.093 | 1      |        |        |        |        |       |
| ENE   | -0.009 | -0.088 | -0.146 | 1      |        |        |        |       |
| INDU  | -0.066 | 0.170  | -0.188 | -0.203 | 1      |        |        |       |
| FDI   | -0.053 | -0.185 | -0.224 | 0.081  | 0.126  | 1      |        |       |
| LNPE  | -0.114 | -0.002 | -0.269 | 0.507  | 0.008  | 0.111  | 1      |       |
| LNEXP | 0.112  | -0.044 | -0.309 | 0.436  | 0.206  | -0.213 | 0.401  | 1     |

**Table 4. Effects of GEA on per capita carbon emissions**

| Variable | RE (1) | FE (2) |
|---|---|---|
| GEA | -1.340* | -1.679** |
|  | (0.721) | (0.771) |
| ER | 117.686 | -23.063 |
|  | (82.633) | (68.850) |
| ENE | 119.894** | 306.763*** |
|  | (56.287) | (75.472) |
| INDU | -1.343* | -1.010 |
|  | (0.779) | (0.954) |
| FDI | -10.538 | -9.802 |
|  | (14.544) | (12.217) |
| LNPE | -0.052 | -0.047 |
|  | (0.040) | (0.033) |
| LNEXP | -3.051*** | -1.909** |
|  | (0.951) | (0.754) |
| Constant | 15.528*** | 5.982** |
|  | (3.785) | (2.871) |
| Province effects | Yes | Yes |
| Time effects | Yes | Yes |
| R-squared | 0.441 | 0.496 |
| Observations | 390 | 390 |

Note: The prefix "LN" before the explanatory variables indicates that the logarithmic form was used. *p < 0.1, **p < 0.05, ***p < 0.01. The numbers in parentheses are the robust standard errors.

According to Column (2) in Table 4, the regression coefficient of the independent variable GEA was negative at the 5% level of significance after controlling for provincial effects, time effects, and the control variables, indicating that GEA has an effect on reductions in carbon reduction effect. Moreover, the results of the estimation with the random effects model in Column (1) corroborated this conclusion; therefore, Hypothesis 1 was confirmed. At the level of the control variables, energy consumption and financial expenditure on environmental protection both had a significant effect on carbon emissions in Columns (1) and (2), with energy consumption significantly and positively correlated with carbon emissions per capita, probably because regions with high electricity consumption tend to have more developed industries and thus higher carbon emissions. However, financial expenditure on environmental protection was significantly and negatively correlated with carbon emissions per capita, reflecting the important role of the government's "visible hand" in reducing carbon emissions.

### 4.3 Robustness test

However, there may be a bidirectional causal relationship between GEA as the independent variable and the dependent variable (carbon emissions), and this may lead to biased results. Therefore, this study referred to the study by Huang et al. (2022), which selects the one-period lag of GEA as the instrumental variable and used two-stage least squares (2SLS) to obtain more accurate findings.

The estimation results for 2SLS are given in Columns (1) and (2) of Table 5, where a significant correlation was found between the instrumental variable L.GEA and GEA. Moreover, the one-stage F-value of the model = 11.6334 > 10, so there was no problem of a weak instrumental variable (Stock & Yogo 2002). In addition, the current period's disturbance term barely affected the estimates of the lags of the GEA index, and thus the instrumental variables also satisfied the assumption of exogeneity. As can be seen in Column (2), even when the instrumental variables were applied to control for the problem of endogeneity, the GEA was still significantly negative at the 10% level and the results of the benchmark regression were again validated.

Table 5. Robustness test

| Variable | 2SLS (1) | 2SLS (2) | Delete municipalities (3) | Replacement of variables (4) |
|---|---|---|---|---|
| GEA |  | -8.229* (4.719) | -1.416* (0.722) |  |
| L.GEA | 0.212*** (0.063) |  |  |  |
| NEW_GEA |  |  |  | -0.023** (0.011) |
| Constant | 0.361* (0.217) | 14.131*** (5.052) | 5.325 (3.285) | 6.103** (2.876) |
| Control | Yes | Yes | Yes | Yes |
| Province effects | Yes | Yes | Yes | Yes |
| Time effects | Yes | Yes | Yes | Yes |
| R-squared | 0.572 | 0.930 | 0.517 | 0.496 |

| Observations | 360 | 360 | 338 | 390 |
|---|---|---|---|---|

Note: *p < 0.1, **p < 0.05, ***p < 0.01. The numbers in parentheses are the robust standard errors.

After dealing with the endogeneity issue, this study also performed a robustness check on the main effects by excluding municipalities and replacing the explanatory variables. The results are shown in Columns (3) and (4) of Table 5. Specifically, Column (3) shows the results after excluding the four municipalities of Beijing, Shanghai, Tianjin, and Chongqing, while Column (4) re-estimates the impact of GEA by changing the core explanatory variable measure, namely using the total number of words instead of the share of word frequency. The results showed that although the GEA and its proxy variables in the different models differed slightly in their estimates and significance levels, they all had a significant inhibitory effect on carbon emissions. Therefore, the negative effect of GEA on carbon emissions can be considered to be robust, and the conclusions drawn on this basis are relatively reliable.

**4.4 Additional analysis**

To explore whether the inhibitory effect of GEA on carbon emissions was regionally heterogeneous, this study referred to the regional classification of the National Bureau of Statistics and the study by Liu et al. (2021). On the basis of the differences in economic development, geographical location, and the demographic structure, China was divided into eastern, central, and western regions (Tibet, Hong Kong, Macau, and Taiwan were excluded because of the availability of data), and a further heterogeneity analysis was conducted.

As shown in Columns (1) to (3) of Table 6, there were regional differences in the impact of GEA on carbon emissions in different regions of China. While the carbon-reducing effect of GEA in the eastern and central regions was not significant, the negative effect of GEA on carbon emissions in the western region was negative at the 5% significance level, and the coefficient was much larger than that estimated in the baseline regression. This shows that the governments in the western region are paying sufficient attention to environmental issues while promoting the transfer of industries from the central and eastern regions. The results also reflect how the competition model among local governments in China is gradually shifting from a single GDP-based competition to a diversified competition.

In addition, to verify the moderating effect of the informatization level on the relationship between GEA and carbon emissions, this study first centralized GEA and the informatization level, and then included them in the model in the form of interaction terms. The resulting moderating effect is presented in the form of a visualization. As can be seen from Figure 3, the negative slope of the straight line was steeper at the high level of informatization, indicating that the higher the level of informatization, the stronger the inhibiting effect of GEA on per capita carbon emissions. The informatization level played a positive moderating role in the relationship between GEA and carbon emissions. Column (4) of Table 6 also supports this conclusion and Hypothesis 2 can be verified. In the era of big data, the synergy between the informatization level and environmental attention is leading to the early achievement of carbon neutrality.

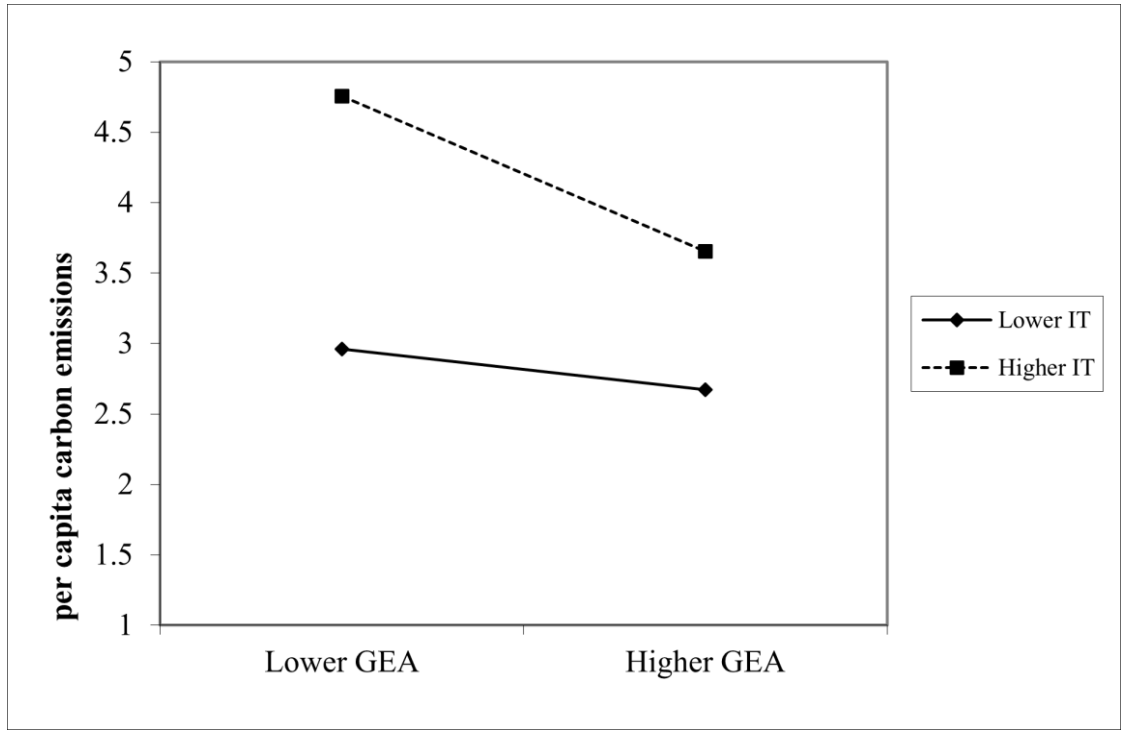

**Figure 3. The moderating effect of informatization level**

**Table 6. Regional heterogeneity and moderating effects**

| Variable | Eastern region (1) | Central region (2) | West region (3) | Full sample (4) |
|---|---|---|---|---|
| GEA | 0.011 | -1.900 | -3.518** | -1.671** |
|  | (0.646) | (1.700) | (1.535) | (0.755) |
| C_GEA×C_IT |  |  |  | -25.956* |
|  |  |  |  | (14.199) |
| IT |  |  |  | 18.477 |
|  |  |  |  | (18.474) |
| Constant | 7.610** | 8.315 | 11.171 | 3.511 |
|  | (3.380) | (6.049) | (10.403) | (3.087) |
| Control | Yes | Yes | Yes | Yes |
| Province effects | Yes | Yes | Yes | Yes |
| Time effects | Yes | Yes | Yes | Yes |
| R-squared | 0.641 | 0.471 | 0.621 | 0.507 |
| Observations | 143 | 104 | 143 | 390 |

Note: *p < 0.1, **p < 0.05, ***p < 0.01. The numbers in parentheses are the robust standard errors.

Considering that the effect of GEA may be non-linear, a panel threshold regression model with GEA as the independent variable and the threshold variable was constructed. Figure 4 shows a plot of the likelihood ratio function for the two thresholds of 0.5474 and 0.5499. From Table 7, it can be seen that the effect of GEA on carbon emissions is indeed non-linear and Hypothesis 2 is tested. However, it is important to note that GEA has a carbon reduction effect in either interval. Also, this effect is magnified to a maximum when the government's attention to the environment is in a reasonable interval (0.5474 < GEA ≤ 0.5499).

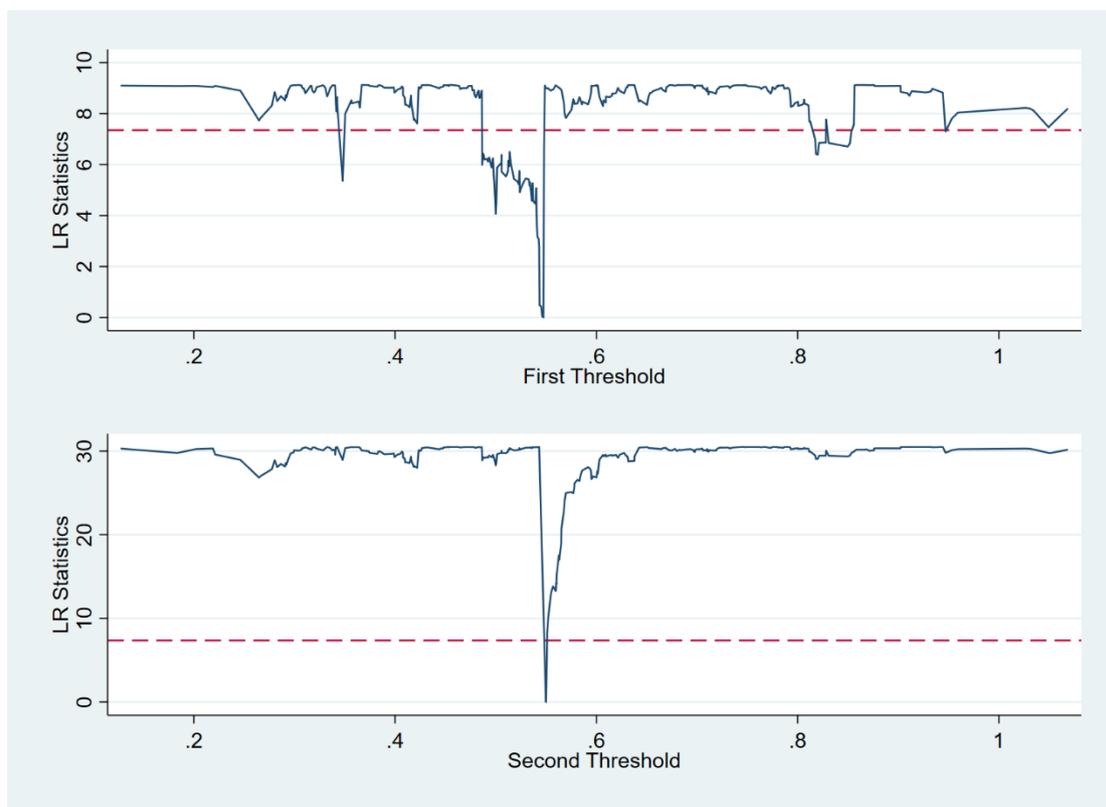

**Figure 4. Likelihood ratio statistics**

**Table 7. Threshold regression results**

| Threshold variable | Regression coefficient | T value |
|---|---|---|
| GEA (GEA≤0.5474) | -2.266* | -1.88 |
| GEA (0.5474＜GEA≤0.5499) | -15.453*** | -6.90 |
| GEA (GEA＞0.5499) | -1.762** | -2.24 |

Note: *p < 0.1, **p < 0.05, ***p < 0.01.

## 5 Conclusions and Policy implications

By using panel data for 30 Chinese provinces from 2007 to 2019, this study examined the impact of governments' environmental attention on per capita carbon emissions and its regional heterogeneity, and analyzed the moderating role of the informatization level in the relationship between governments' environmental attention and per capita carbon emissions. The main findings are as follows.

(1) From 2007 to 2019, the environmental attention index of local governments in China showed an overall fluctuating upward trend, but there was still much room for improvement. (2) Governments' environmental attention had a significant carbon-reducing effect, and this finding still held after the test of robustness. (3) The results of the heterogeneity analysis showed that the carbon-reducing effect of the governments' environmental attention was regionally heterogeneous; specifically, the carbon-reducing effect was more significant in western China but it was not significant in central and eastern China. (4) There is a synergistic effect between the governments' environmental attention and the informatization level, and the synergy between the two has a significant contribution to reducing carbon emissions. (5) There is a significant threshold effect on the carbon reduction effect of governments' environmental attention.

In light of these findings, this study proposes the following policy implications: While anxiety about climate change and concern about environmental governance is driving a shift in the focus of governmental attention, there are still fluctuations in governmental attention to the environment that need to be improved. Therefore, the central government should continue to emphasize the importance of ecological civilization in The Five-Sphere Integrated Plan, while striving to explore the best way to embed mobilized governance in environmental issues, such as environmental protection inspectors, and further build a normalized mechanism for mobilized governance and, ultimately achieve a sustainable improvement in the governments' environmental attention. Secondly, the institution of incentive and constraint functions could be brought into play to promote the transformation of the governments' environmental attention into action towards environmental governance. On the one hand, it can be used to incentivize local governmental decision-makers to actively take action to reduce carbon emissions; on the other hand, it could be used to restrain the failure to implement policies brought about by the transfer of officials. Thirdly, the more significant carbon-reducing effect of the governments' environmental attention in western China also suggests that the western region should give more prominence to ecological benefits while taking over the transfer of industry from eastern and central China, and reasonably allocate the proportion of attention and resources between economic development and ecological environment to promote the green transformation of local economic development. Finally, the synergy between the informatization level and the governments' environmental attention suggests that public participation, expert discussion, and media focus should be strengthened in the agenda of policies for reducing carbon. Local governments also need to continue to promote the disclosure of government information to adapt to the trend of digital government, and to use disclose information to achieve synergistic governance between the public and the government in actions to reduce carbon reduction.

# Declarations

**Conflict of interest** The author declares that there are no competing interests.

206